\documentclass[12pt]{article}
\usepackage{epsf}
\usepackage{feynmf}
\begin{fmffile}{dydiagram}

\newcommand{\ie}{{\it i.e.}}

\newcommand{\re}{{\rm Re}}
\newcommand{\qu}{{\rm q}}

\newcommand{\qbm}{{\rm\bar q}}

\newcommand{\qvec}{\vec q}
\newcommand{\pvec}{\vec p}
\newcommand{\kvec}{\vec k}
\newcommand{\rvec}{\vec r}
\newcommand{\Rvec}{\vec R}

\newcommand{\M}{{\cal M}}
\newcommand{\pl}{{||}}

\newcommand{\order}[1]{${\cal O}\left(#1 \right)$}
\newcommand{\morder}[1]{{\cal O}\left(#1 \right)}
\newcommand{\eq}[1]{(\ref{#1})}

\newcommand{\ave}[1]{\langle{#1}\rangle}

\newcommand{\beq}{\begin{equation}}
\newcommand{\eeq}{\end{equation}}
\newcommand{\beqa}{\begin{eqnarray}}
\newcommand{\eeqa}{\end{eqnarray}}

\renewcommand{\thefootnote}{\fnsymbol{footnote}}

\begin{document}

\begin{titlepage}
\begin{flushright}
           19 July 2002\\
           LAPTH-917/02\\
           hep-ph/0206138\\
\end{flushright}

\vskip 1.5cm

\centerline{\bf ABSENCE OF SHADOWING IN DRELL-YAN PRODUCTION}
\vskip 1mm
\centerline{\bf AT FINITE TRANSVERSE MOMENTUM EXCHANGE}

\vskip 1.cm

\centerline{St\'ephane Peign\'e}

\vskip .5cm

{\small\sl
\centerline{LAPTH\footnote{CNRS, UMR 5108, associated to the
      University of Savoie.}, BP 110, F-74941 Annecy-le-Vieux Cedex, France}}

\vskip 1cm

\begin{abstract} Within a perturbative scalar QED model
recently considered by Brodsky et al.,
we study how leading-twist Coulomb rescatterings affect 
the Drell-Yan cross section
at small $x = x_{target}$, and compare to the case of 
deep inelastic scattering 
at small $x_{B}$. We show that in the range where the 
transverse momentum transferred to the target 
is large compared to its minimal value $\sim \morder{x}$, Coulomb 
rescatterings affect the DIS cross section but not the 
Drell-Yan production rate. This illustrates that the 
leading-twist parton distribution functions become non-universal
when cross sections which are differential in target-related
particles are considered. 
\end{abstract}

\end{titlepage}

\newpage
\renewcommand{\thefootnote}{\arabic{footnote}}
\setcounter{footnote}{0}
\setcounter{page}{1}

{\bf \Large 1. Introduction and summary}
\vskip 5mm

Within the parton model, deep inelastic lepton-nucleon scattering 
structure functions have been shown to measure the probability
to find in the target nucleon a parton with longitudinal momentum
fraction $x_{Bjorken}= x_B$ in the infinite momentum frame \cite{dy}.
This result was obtained in a theory of pions and nucleons for the 
strong interaction. Since, the correct theory of the strong
interaction has been established to be a gauge theory, QCD. According
to QCD factorization theorems \cite{fact}, at leading-twist the inclusive
deep inelastic scattering (DIS) and Drell-Yan (DY) cross sections
(in particular) can be factorized and expressed as convolutions
between quark and gluon distributions in the incoming hadron(s) and
the partonic subprocess cross sections. The predictive power of
factorization theorems arises from the statement that parton
distribution functions are {\it universal} quantities, \ie\
independent of the collision. The universality of parton
distributions appears to be supported by the data, at least up to some
accuracy. 
Also, the quark distribution (in the nucleon
$N$ of momentum $p$) probed in DIS,
\beqa
f_{\qu/N}(x_B,Q^2)&=& \frac{1}{8\pi} \int dy^- \exp(-ix_B p^+ y^-)
\label{fq}\\
&\times&\langle N(p)| \qbm(y^-) \gamma^+\, {\rm P}\exp\left[ig\int_0^{y^-}
dw^- A^+(w^-) \right] \qu(0)|N(p)\rangle \nonumber
\eeqa
where all fields are evaluated at equal light-cone time $y^+=0$
and transverse position $\vec{y}_{\perp} = \vec{0}_{\perp}$,
seems directly related to the nucleon light-cone wavefunction in 
$A^+=0$ gauge, supporting the probabilistic interpretation of the 
parton distribution functions (and hence of the DIS structure
functions), as in the original parton model.

But the expression \eq{fq} is incorrect in $A^+=0$ gauge, \ie\ the 
quark distribution is not given by the (squared) nucleon light-cone
wavefunction \cite{bhmps}. 
Roughly speaking, this is because in the Bjorken $\nu \to \infty$
limit, the eikonal 
coupling of the struck quark of momentum $p_1$ to the target color
field $A^{\mu}$ satisfies 
$p_1 \cdot A \propto \nu A^+ \to \infty$ in all gauges,
{\it except} $A^+=0$. More precisely, 
although the light-cone time $y^+$ between the
absorption and emission 
of the virtual photon in the forward DIS amplitude vanishes as
$1/\nu$, Coulomb interactions occurring in this short
time interval actually modify the DIS cross section at leading-twist
in all gauges, including the light-cone $A^+=0$ gauge \cite{bhmps}. 
Thus in a {\it gauge} theory,
the simple identification between parton distribution and parton
probability (defined as the square of the nucleon light-cone
wavefunction) does not hold. Although not excluded by this
observation, the universality of parton distributions becomes much
less intuitive. In this respect it was recently shown that
single transverse spin asymmetries in semi-inclusive DIS appear at
leading-twist \cite{ssabhs}, correcting previous statements 
\cite{0ssacollins}. This is due to the
non-universality of spin-dependent parton distributions (in other
words of the Sivers asymmetry \cite{sivers}), which originates from
the subtle behaviour of Wilson lines under time-reversal
\cite{ssacollins}. 
A possible correct expression in light-cone gauge 
for the gauge link entering the definition of
spin-dependent parton distributions has recently been suggested
\cite{jy}.

In this context, it is important to reconsider the question 
of universality of spin-independent parton distributions.
In the present work, I compare in a simple model the spin-independent quark
distributions probed in DIS and in the Drell-Yan process at small
values of $x$. I show that in the range of the exchanged transverse
momentum $k_{\perp}$ responsible for leading-twist shadowing in DIS,
the Coulomb rescattering corrections a priori modifying the DY cross
section are in fact unitary. This is similar to what Bethe and Maximon 
found in the case of high energy bremsstrahlung and pair production
\cite{bm}. Before the advent of QCD, it was also found that
corrections to the parton model Drell-Yan formula are actually absent
\cite{cw,del}. In the context of gauge theories, our result is an
example of the non-universality of the leading-twist parton
distributions, which arises when considering a cross section which is
{\it differential} in the target structure.
However, according to the QCD factorization theorem for 
{\it inclusive} cross sections, 
we expect the {\it $k_{\perp}$-integrated} quark
distributions probed in DIS and DY to be identical, even though the
typical $k_{\perp}$ contributing in both cases is different. 
We check in Appendix A that this identity indeed occurs within a model
where the scale $\sim \morder{x}$ is screened by a finite photon mass.
But the fact that this holds in general, for any target, is not obvious, 
and we think further studies are needed to settle (or disprove) 
the universality of parton distributions.

I briefly review in section 2 the model of Brodsky et al. 
developed in Ref.~\cite{bhmps} 
for DIS shadowing at small $x_B$. We recall that this model
concentrates on the {\it leading-twist} shadowing correction to the
DIS cross section (arising from the aligned-jet kinematic region),
which can be interpreted as part of the target quark
distribution function probed in DIS. The typical value 
$\ave{k_{\perp}}_{DIS}$ of the exchanged transverse momentum is
found to be of the order of a soft but 
$x_B$-independent scale, $\ave{k_{\perp}}_{DIS} \sim \morder{m}$.
The model is simply extended to DY production in section 3. 
Similarly to DIS, the leading-twist Coulomb corrections to the DY
cross section arise from a kinematic region 
which we call the `aligned-photon' region 
(by analogy with the aligned-jet region of DIS) where the longitudinal
momentum fraction taken from the incoming projectile (anti)quark by the
radiated virtual photon approaches unity. Those corrections are
interpreted as part of the quark distribution probed in the DY
process.  
We find that for $Mx \ll m$, where $M$ is the target mass and 
$x = x_{target} \ll 1$, 
the DY cross section is unaffected by Coulomb
rescattering at values $k_{\perp} \sim \morder{m}$, contrary to the 
DIS cross section. This is the main result of the present paper.

This result is obtained in a scalar QED model and in the limit 
$x \ll 1$, which allows great technical simplications in the loop
calculations. Since we neglect the scale $Mx$ compared to $k_{\perp}$ 
from the beginning, the $k_{\perp}$-integrated DY cross
section is out of reach in the present model. Thus we cannot exclude 
that the {\it total} DY cross section receives a non-zero leading-twist
shadowing correction. 
However, if this happens, the typical value of
$k_{\perp}$ responsible for this effect must be, for $x \ll 1$,
\beq
\label{dydis}
\ave{k_{\perp}}_{DY}  \sim Mx \ll \ave{k_{\perp}}_{DIS} 
\eeq
This might have some implications on the properties of momentum 
broa\-dening and energy loss in the Drell-Yan process. We note that 
the observed difference between the nuclear broadening of the average
transverse momentum in DY production and in dijet photoproduction
is not understood \cite{guo}. The result \eq{dydis} might give some
hint to this problem. 

But more importantly, it might question
the universality of parton distributions at small $x$, 
as we will discuss in section 4. In this respect, let us note that our
result, namely the fact that Coulomb rescatterings do not modify the
leading-twist DY Born cross section in the region of transverse
momentum exchange $k_{\perp} \gg Mx$, is similar to what was found in
Ref.~\cite{lrs}. There it was shown, for transverse momenta being large
compared to infrared cut-offs (and much smaller than the collision
energy), that long-distance contributions to the DY cross section
cancel out at the two-loop order. This was argued to be a good
indication for the validity of factorization. I stress that it makes
on the contrary factorization much less evident, 
since in the same transverse
momentum domain, Coulomb rescatterings {\it modify} the DIS cross
section, resulting in the observed nuclear shadowing of the DIS parton
distributions. 


\vskip 1cm
{\bf \Large 2. Leading-twist shadowing in DIS}
\vskip 5mm

{\bf \large 2.1 Model for the quark distribution function}

A perturbative model for leading-twist DIS shadowing
has recently been studied in \cite{bhmps}. Before extending this model 
to the DY 
process in the next section, we recall its main features.
A specific contribution to $\sigma_{DIS}$ is evaluated, via the 
optical theorem, from the forward DIS amplitude shown in Fig.~1.

\begin{figure}[htb]
\begin{center}
\leavevmode
{\epsfxsize=10truecm \epsfbox{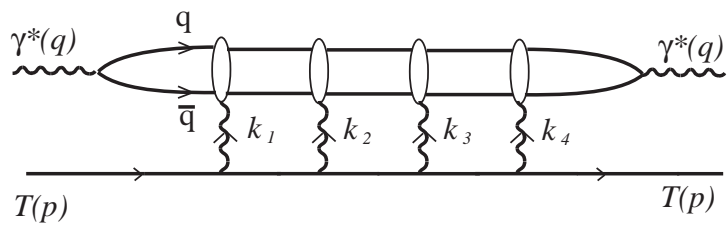}}
\end{center}
\caption[*]{Forward $\gamma^* T \to \gamma^* T$ amplitude in the DIS
  model of Ref.~\cite{bhmps}.}
\label{fig1dy}
\end{figure}

The model is perturbative and chosen to be scalar QED. One takes
for the target a scalar ``quark'' of mass $M$ and momentum $p$, and for the 
light ``quark'' and ``antiquark'' scalars of mass $m$ and momenta 
$p_1$ and $p_2$. The couplings of the ``gluons'' of momenta $k_i$ 
and of the incoming virtual photon of momentum $q$ to the scalars are
denoted by $g$ and $e$ respectively.
The forward amplitude of Fig.~1 contributes to $\sigma_{DIS}$ through three
different cuts between the Coulomb gluon exchanges. Calling $A$, $B$ 
and $C$ the single, double, and three-gluon exchange amplitudes 
for the process
$\gamma^*(q) T(p) \to \qu(p_1)\qbm(p_2) T(p')$, the rescattering correction 
of order $e^2 g^8$ to the Born term $\int |A|^2$ reads
\beq
\label{deltasigmaDIS}
\Delta \sigma_{DIS}  \sim \int
\frac{d^2\pvec_{2\perp}}{(2\pi)^2}\, \frac{d^2\kvec_\perp}{(2\pi)^2}\,
\left[ |B|^2 + 2 \re(A^* C) \right]
\eeq
Feynman diagrams contributing to $A$ are shown in Fig.~2. 

\begin{figure}[htb]
\begin{center}
\leavevmode
{\epsfxsize=12truecm \epsfbox{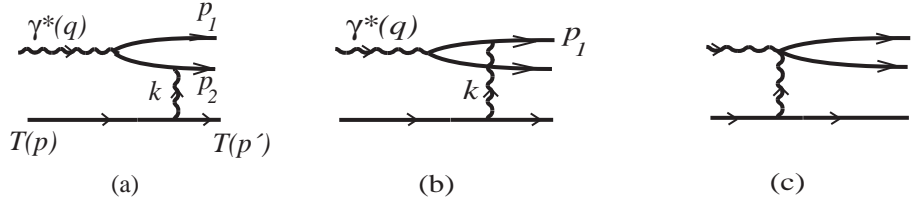}}
\end{center}
\caption[*]{Single gluon exchange DIS amplitude in scalar QED.}
\label{fig2dy}
\end{figure}

The amplitudes $B$ and $C$ are obtained by adding to the Born
amplitude $A$ one or two gluon exchanges between the target and the
light quarks. 
In the Bjorken limit\footnote{The Bjorken limit
is defined as $q^- = 2\nu \to \infty$, $Q^2 = -q^2 \to \infty$ with
$x_B = \frac{Q^2}{2M\nu}$ being fixed. We use the light-cone variables
$q^\pm = q^0 \pm q^z$.} and at small $x_B$, $\Delta \sigma_{DIS}$ receives
a leading-twist contribution, arising from the aligned-jet
configuration and presenting the features of a shadowing correction 
to the DIS Born cross section \cite{bhmps}. It was shown that the 
kinematic region where leading-twist shadowing appears reads:
\beq
2\nu \sim p_1^- \gg p_2^- \gg k_{\perp},\ k_{i\perp},\ p_{2\perp},
\ k_{i}^-, \ m,\ M \gg k_i^+,\ k^+,\ p_2^+ \sim \morder{Mx_B}
\label{MscalesDIS}
\eeq
where when $\nu \to \infty$
the total momentum transfer $k$ satisfies
\beq
\label{kplus}
k^+ = Mx_B+ p_2^+
\eeq

The kinematic limit \eq{MscalesDIS} holds in the target rest frame,
where in the four-momentum notation $k=(k^+, k^-, \kvec_{\perp})$ we
have:
\beqa 
\label{4vectorsDIS}
p &=& (M, M, \vec{0}_{\perp}) \nonumber \\
q &=& (-Mx_B, 2\nu, \vec{0}_{\perp}) \nonumber \\
\epsilon_L &=& \frac{Q}{\nu}(1, -1, \vec{0}_{\perp})
\eeqa
In the case of scalar QED, the leading-twist contribution to 
$\sigma_{DIS}$ arises from the light quarks coupling 
to a photon with longitudinal polarization $\epsilon_L$.

The scale $\nu$ is the single hard scale in the problem,
and the limit $\nu \to \infty$ is taken from the beginning.
In the aligned-jet kinematics $q^- = 2\nu \simeq p_1^-$, Coulomb
rescattering corrections contribute at leading-twist to the DIS cross
section \cite{pw}. Compared to the scale $\nu$, the antiquark has a soft
momentum $p_2$ and  
must be considered as part of the (soft) target dynamics \cite{fact}.
(At small $x_B$, $p_2^- \propto 1/x_B$ can however become large 
enough, so that the physics of destructive interferences between
diffractive amplitudes takes place, resulting in shadowing).
In addition, the hard vertex $\gamma^* \qu \to \qu$ (as viewed in the
infinite momentum frame) is taken 
at zeroth order in the strong coupling $g$.
Hence the contribution to $\Delta \sigma_{DIS}$ arising from 
the domain \eq{MscalesDIS} is a perturbative model
for the scaling target quark distribution $f_{\qu/T}(x_B)$.
The leading-twist contribution to $\Delta \sigma_{DIS}$ found
in \cite{bhmps} is thus interpreted as shadowing of the quark distribution
function in the target.

In order to compare the quark distributions probed in DIS and DY, 
we will apply the model described above to the DY process
in the next section. 
Let us repeat before the results obtained in \cite{bhmps}
for the DIS amplitudes $A$, $B$, $C$ and for the shadowing 
correction $\Delta \sigma_{DIS}$.

\vskip 5mm

{\bf \large 2.2 DIS rescattering amplitudes}

{\bf Born amplitude}

At small $x_B$ the Born amplitude for the DIS process is obtained 
in Feynman gauge from the dominant diagrams of Figs.~2a
and 2b, and in light-cone $A^+=0$ gauge from the diagram of Fig.~2a
only. The gauge invariant result reads in momentum space:
\beq A(p_2^-,\pvec_{2\perp},\kvec_\perp) =
\frac{2eg^2 M Q p_2^-}{k_\perp^2}\left[\frac{1}{D(\pvec_{2\perp})}
- \frac{1}{D(\pvec_{2\perp}-\kvec_\perp)} \right]
\label{ADIS}
\eeq
where
\beqa
D(\pvec_{\perp}) &=& p_{\perp}^2 + m_\pl^2 \label{Ddef}\\
m_\pl^2 &=& p_2^-Mx_B + m^2 \label{mpl} 
\eeqa 
In transverse coordinate space we have
\beqa
\tilde A(p_2^-,\rvec_\perp, \Rvec_\perp) &=& \int
\frac{d^2\pvec_{2\perp}}{(2\pi)^2}\, \frac{d^2\kvec_\perp}{(2\pi)^2}\,
A(p_2^-,\pvec_{2\perp},\kvec_\perp) \exp\left(i\rvec_\perp\cdot\pvec_{2\perp}
+ i\Rvec_\perp\cdot\kvec_{\perp} \right) \nonumber \\
&=& 2eg^2 M Q p_2^-\, V(m_\pl r_\perp) W(\rvec_\perp, \Rvec_\perp)
\label{AtildeDIS}
\eeqa
The functions $V$ and $W$ stand respectively for the incoming photon
wavefunction describing its $\qu \qbm$ content and for the $\qu \qbm$
dipole scattering amplitude:
\beqa
V(m\, r_\perp) &\equiv&
\int \frac{d^2\pvec_\perp}{(2\pi)^2}
\frac{e^{i\rvec_\perp\cdot\pvec_{\perp}}}{p_\perp^2+m^2}
= \frac{1}{2\pi}K_0(m\,r_\perp)
\label{Vexpr} \\
W(\rvec_\perp, \Rvec_\perp) &\equiv&
\int \frac{d^2\kvec_\perp}{(2\pi)^2}
\frac{1-e^{i\rvec_\perp\cdot\kvec_{\perp}}}{k_\perp^2}
e^{i\Rvec_\perp\cdot\kvec_{\perp}} = \frac{1}{2\pi}
\log\left(\frac{|\Rvec_\perp+\rvec_\perp|}{R_\perp} \right)
\label{Wexpr}
\eeqa

{\bf Two-gluon exchange}

The gauge invariant expression of the one-loop DIS amplitude $B$
cor\-res\-pon\-ding to two gluon exchanges 
between the target and the light quarks is \cite{bhmps}:
\beqa
&& B(p_2^-,\pvec_{2\perp},\kvec_\perp) = -ieg^4 M Q p_2^- \int \frac{d^2
\kvec_{1\perp}}{(2\pi)^2}\ \frac{1}{k_{1\perp}^2 k_{2\perp}^2}
\label{BDIS} \nonumber\\
&\times& \left[
\frac{1}{D(\pvec_{2\perp})} - \frac{1}{D(\pvec_{2\perp}-\kvec_{1\perp})}
-\frac{1}{D(\pvec_{2\perp}-\kvec_{2\perp})}+
\frac{1}{D(\pvec_{2\perp}-\kvec_{\perp})} \right] 
\eeqa
where $\kvec_{2\perp} = \kvec_{\perp} - \kvec_{1\perp}$.
In transverse coordinate space:
\beq 
\tilde B(p_2^-,\rvec_\perp, \Rvec_\perp) = -ieg^4 M Q p_2^- V(m_\pl
r_\perp) W^2(\rvec_\perp, \Rvec_\perp)= \frac{-ig^2}{2!} W \tilde
A \label{BtildeDIS} 
\eeq

{\bf Three-gluon exchange}

We give the expression of the three-gluon exchange amplitude
$C$ found in \cite{bhmps}:
\beqa
&& C(p_2^-,\pvec_{2\perp},\kvec_\perp) = -\frac{1}{3} eg^6 M Q p_2^-\int
\frac{d^2 \kvec_{1\perp}}{(2\pi)^2}
\frac{d^2 \kvec_{2\perp}}{(2\pi)^2}\ \frac{1}{k_{1\perp}^2\,
k_{2\perp}^2\, k_{3\perp}^2} 
\label{CDIS} \nonumber\\
&\times& \left[ \frac{1}{D(\pvec_{2\perp})} -\frac{3}{D(\pvec_{2\perp}
-\kvec_{1\perp})} +
\frac{3}{D(\pvec_{2\perp}- \kvec_{1\perp}-\kvec_{2\perp})} -
\frac{1}{D(\pvec_{2\perp} -\kvec_{\perp})} \right] \nonumber\\
\eeqa
where $\kvec_{3\perp}=\kvec_{\perp}- \kvec_{1\perp}- \kvec_{2\perp}$.
In coordinate space:
\beq
\tilde C(p_2^-,\rvec_\perp, \Rvec_\perp) = -\frac{1}{3}eg^6 M Q p_2^-
V(m_\pl r_\perp) W^3(\rvec_\perp, \Rvec_\perp)=
\frac{(-ig^2)^2}{3!}W^2 \tilde A
\label{CtildeDIS}
\eeq

\vskip 5mm
{\bf \large 2.3 The $k_{\perp}$-range in DIS}

We stress here that the amplitudes $B$ and $C$ are infrared finite.
This is because the quark $p_1$ and antiquark $p_2$ form a dipole,
whose scattering amplitude $W$ vanishes with the separation $r_{\perp}$
between the two quarks (see \eq{Wexpr}). Thus in \eq{BDIS} and 
\eq{CDIS} the typical values of $k_{i\perp}$ are 
$\sim \morder{k_{\perp}, p_{2\perp}}$. The only other (soft) scale
present being
$m_\pl$ given in \eq{mpl}, the typical value of the total exchanged 
transverse momentum $k_{\perp}$ contributing to the 
$k_{\perp}$-integrated correction $\Delta \sigma_{DIS}$ is:
\beq
\label{kperpscaleDIS}
\ave{k_{\perp}}_{DIS} \sim m_\pl \sim \morder{m}
\eeq
The rescattering 
correction can be obtained from
Eqs.~\eq{BtildeDIS} and \eq{CtildeDIS}:
\beq
\Delta \sigma_{DIS} \sim \int d^2\rvec_\perp\, d^2\Rvec_\perp\,
\left[ |\tilde B|^2 + 2 {\tilde A} {\tilde C} \right] = 
- \frac{1}{3} \int d^2\rvec_\perp\, d^2\Rvec_\perp \, 
\frac{g^4}{4} {\tilde A}^2 W^2 
\label{sigmaDIS}
\eeq
This is the leading-twist shadowing correction to the Born DIS cross
section found in \cite{bhmps}, interpreted as part of the (scalar) 
quark distribution $f_{\qu/T}(x_B)$.

\vskip 1cm
{\bf \Large 3. Rescattering effects in Drell-Yan production}
\vskip 5mm

{\bf \large 3.1 Model for Drell-Yan production}

We now extend the model presented previously for DIS to
the Drell-Yan process. This can be done 
by simply exchanging the virtual photon $q$ and the 
quark $p_1$. We thus describe DY production in the target rest frame
where the incoming antiquark has a large `minus' momentum component,
$p_1^- \simeq 2\nu$. 
As we will see the basic process for DY production in this frame
corresponds to quark-antiquark annihilation in the infinite momentum
frame. 

One diagram contributing to the DY forward amplitude is
represented in Fig.~3. All diagrams are simply obtained by
taking into account all possible permutations of the lower and
upper vertices. 

\vskip 1.5cm 

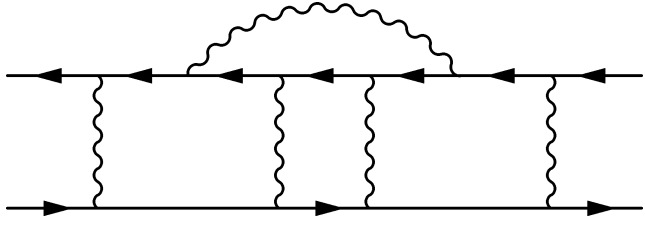
\begin{figure}[htb]
\begin{center}
\leavevmode
\begin{fmfgraph*}(300,50)
   \fmfleft{v1,v12}
   \fmfright{v6,v7}
   \fmf{fermion,label=$p$,label.side=right}{v1,v2}
   \fmf{plain}{v2,i,v3}
   \fmf{fermion,label=$p'$,label.side=right}{v3,v4}
   \fmf{plain}{v4,j,v5}
   \fmf{fermion,label=$p$,label.side=right}{v5,v6}
   \fmf{fermion,label=$p_1$,label.side=right}{v7,v8}
   \fmf{fermion}{v8,k,v9}
   \fmf{fermion,label=$p_2$,label.side=right}{v9,v10}
   \fmf{fermion}{v10,l,v11}
   \fmf{fermion,label=$p_1$,label.side=right}{v11,v12}
   \fmffreeze
   \fmf{photon,label=$k_1$,label.side=left}{v2,v11}
   \fmf{photon,label=$k_2$,label.side=left}{v3,v10}
   \fmf{photon,label=$k_3$,label.side=right}{v4,v9}
   \fmf{photon,label=$k_4$,label.side=right}{v5,v8}
   \fmf{photon,left=.5,label=$q$,label.side=left}{l,k}
\end{fmfgraph*}
\end{center}
\caption[*]{Forward amplitude of order $e^2 g^8$ for the Drell-Yan
   process. Only one diagram is shown.} 
\label{fig3dy}
\end{figure}

The Born DY cross section will get a rescattering correction:
\beq
\label{deltasigmaDY}
\Delta \sigma_{DY}  \sim \int
\frac{d^2\pvec_{2\perp}}{(2\pi)^2}\, \frac{d^2\kvec_\perp}{(2\pi)^2}\,
\left[ |B_{DY}|^2 + 2 \re(A_{DY}^* C_{DY}) \right]
\eeq
where $A_{DY}$, $B_{DY}$, $C_{DY}$ are the amplitudes for the 
process $\qbm(p_1)T(p)\to\gamma^*(q)\qbm(p_2)T(p')$ 
corresponding to 
one, two, and three-gluon exchange. In the following we will evaluate 
these amplitudes in the small $x$ limit.

In the present DY case the photon momentum $q$ is time-like,
$q^2 = Q^2 >0$ is the final lepton pair invariant mass squared, 
and the momenta are chosen as ($q^+ >0$):
\beqa
q &=& (+M x, q^-, \qvec_{\perp}) \nonumber \\
p_1 &=& (p_1^+, 2 \nu, \vec{0}_{\perp}) \nonumber \\
p &=& (M, M, \vec{0}_{\perp})
\eeqa
where 
\beq \label{x}
x = \frac{Q^2}{2M\nu}
\eeq
It is easy to check that the configuration 
$p_1^- = 2\nu \simeq q^- \to \infty$, which we call the
`aligned-photon' configuration by analogy to the DIS aligned-jet region,
gives a leading-twist contribution to $\Delta \sigma_{DY}$.
In the DY calculation the same longitudinal
photon polarization vector as for DIS can be used.

In the $\nu \to \infty$ limit the total momentum transfer $k$ 
still satisfies (see \eq{kplus}):
\beq
\label{kplusDY}
k^+ = Mx + p_2^+
\eeq
The relevant kinematics in the target rest frame 
is similar to \eq{MscalesDIS}, 
\beq
2\nu = p_1^- \gg p_2^- \gg k_{\perp},\ k_{i\perp},\ p_{2\perp},\ k_i^-,
\ m ,\ M,\ q_{\perp} \gg k_i^+,\ k^+ = Mx+ p_2^+
\label{MscalesDY}
\eeq
where one just added the soft $q_{\perp}$ scale. 

As in DIS, the antiquark $p_2$ is part 
of the soft target dynamics. The incoming ``hadron'' is modelled as a single 
antiquark, whose energy $\nu$ is transferred totally to the virtual 
photon. Thus in the present model the colliding partons 
from the projectile and  
target carry respectively the momentum fractions 
$x_1 = 1 \simeq x_F$ and $x_2 = (k^+ - p_2^+)/p^+ = x = Q^2/(2M\nu)$.
In the infinite momentum frame, we recover quark-antiquark
annihilation as the basic partonic process for DY production.

The hard $\qu \qbm \to \gamma^{*}$ vertex is still taken at zeroth 
order in $g$, thus all the soft dynamics should be
interpreted as part of the target quark distribution, probed at a value
$x$ of the longitudinal momentum fraction. Since the shadowing contribution 
found in DIS describes the target quark distribution probed at $x_B=x$,
one would naively expect, assuming parton distributions to be universal,
to find a rescattering correction to the DY Born cross section 
originating from the domain \eq{MscalesDY} equal 
to that of DIS.

As we will show, in the region \eq{MscalesDY} the rescattering
corrections to $\sigma_{DY}$ are unitary, \ie, do not modify the Born 
DY cross section, contrary to the DIS case. 
In this sense the effect of shifting the {\it outgoing} quark of DIS 
to an {\it incoming} antiquark in DY is drastic.

\vskip 5mm
{\bf \large 3.2 DY rescattering amplitudes}

I now give the DY amplitudes in the small $x$ limit.  
The calculation has been performed both in Feynman 
and light-cone $A^+=0$ gauge, yielding gauge-invariant results.
Since different diagrams can contribute
in these two gauges, for simplicity the following discussion refers to the 
Feynman gauge calculation.

{\bf Born amplitude}

The Born amplitude for the DY process is given
in Feynman gauge by the diagrams obtained by exchanging
$q$ and $p_1$ in Figs.~2a and 2b. The result in the small $x$ limit
reads:
\beq
 A_{DY}(p_2^-,\pvec_{2\perp},\kvec_\perp) =
- \frac{2eg^2 M Q p_2^-}{k_\perp^2}\left[\frac{1}{D(\pvec_{2\perp})}
- \frac{1}{D(\pvec_{2\perp}-\kvec_\perp)} \right]
\label{ADY}
\eeq
This is equal to the Born amplitude obtained for DIS 
(see \eq{ADIS}), up to an irrelevant sign. 
This sign arises since the coupling of the photon brings a factor
$\epsilon_L \cdot (2p_1-q) =Q$ for the DIS amplitude, and 
$\epsilon_L \cdot (q-2p_1)$ for the DY amplitude. This is due to the fact
we consider for DY an incoming {\it anti}quark of momentum $p_1$.

{\bf Two-gluon exchange}

In Feynman gauge, the one-loop diagrams which dominate in the 
small $x$ limit are shown in Fig.~4. The `crossed' diagrams,
obtained by permuting the photon coupling vertices to the target line,
are also taken into account.
We found that the diagrams where the virtual photon emission
occurs between the two gluon exchanges are suppressed in this limit
(see one example in Fig.~5, where the crossed diagram is also
implicitly included). 
This suppression of radiation in DY 
production has been mentioned previously \cite{bhq}, but we stress here
that it occurs only when the
transferred momenta $k_{i\perp}$ are large compared to 
$Mx$, which is precisely the limit studied here (see \eq{MscalesDY}). 

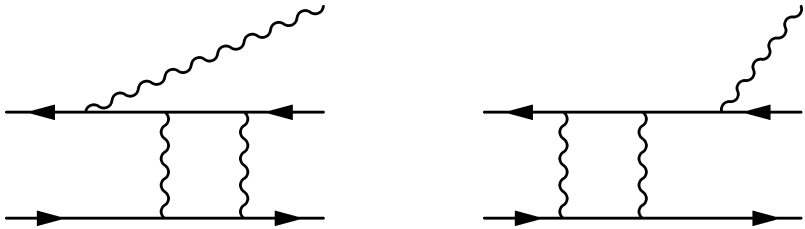
\begin{figure}[t]
\begin{center}
\leavevmode
\begin{fmfgraph*}(120,80)
   \fmfstraight\fmfleft{v1,v8,p}
   \fmfstraight\fmfright{v4,v5,q}
   \fmf{fermion,label=$p$,label.side=right}{v1,i}
   \fmf{plain}{i,v2,v3}
   \fmf{fermion,label=$p'$,label.side=right}{v3,v4}
   \fmf{fermion,label=$p_2$,label.side=right}{v5,v6}
   \fmf{plain}{v6,v7,j}
   \fmf{fermion,label=$p_1$,label.side=right}{j,v8}
   \fmf{photon,tension=0,label=$k_1$,label.side=left}{v2,v7}
   \fmf{photon,tension=0,label=$k_2$,label.side=right}{v3,v6}
   \fmffreeze 
   \fmf{photon,label=$q$,label.side=left}{j,q}
\end{fmfgraph*}
\hskip 2cm 
\begin{fmfgraph*}(120,80)
   \fmfstraight\fmfleft{v1,v8,p}
   \fmfstraight\fmfright{v4,v5,q}
   \fmf{fermion,label=$p$,label.side=right}{v1,v2}
   \fmf{plain}{v2,v3,i}
   \fmf{fermion,label=$p'$,label.side=right}{i,v4}
   \fmf{fermion,label=$p_2$,label.side=right}{v5,j}
   \fmf{plain}{j,v6,v7}
   \fmf{fermion,label=$p_1$,label.side=right}{v7,v8}
   \fmf{photon,tension=0,label=$k_1$,label.side=left}{v2,v7}
   \fmf{photon,tension=0,label=$k_2$,label.side=right}{v3,v6}
   \fmffreeze
   \fmf{photon,label=$q$,label.side=left}{j,q}
\end{fmfgraph*}
\end{center}
\caption[*]{Diagrams contributing dominantly to the two-gluon exchange
   DY amplitude in Feynman gauge and in the small $x$ limit. Crossed
   diagrams, obtained by permuting the lower vertices, are included.}
\label{fig4dy}
\end{figure}

\begin{figure}[t]
\begin{center}
\leavevmode
\begin{fmfgraph*}(200,80)
   \fmfstraight\fmfleft{p,k,p1,i}
   \fmfstraight\fmfright{pp,l,p2,q}
   \fmfstraight\fmfbottom{p,v2,u,v1,pp}
   \fmf{fermion,label=$p$,label.side=right}{p,v2}
   \fmf{plain}{v2,u,v1} 
   \fmf{phantom_arrow,tension=0}{v2,v1}
   \fmflabel{$p-k_1$}{u}
   \fmf{phantom_arrow}{v2,v1}
   \fmf{fermion,label=$p'$,label.side=right}{v1,pp}
   \fmf{fermion,label=$p_2$,label.side=right}{p2,v5}
   \fmf{fermion,label=$p_2-k_2$,label.side=left}{v5,v4}
   \fmf{fermion,label=$p_1+k_1$,label.side=left}{v4,v3}
   \fmf{fermion,label=$p_1$,label.side=right}{v3,p1}
   \fmf{photon,tension=0,label=$k_1$,label.side=left}{v2,v3}
   \fmf{photon,tension=0,label=$k_2$,label.side=right}{v1,v5}
   \fmffreeze
   \fmf{photon,label=$q$,label.side=left}{v4,q}
\end{fmfgraph*}
\end{center}
\caption[*]{A diagram (together with the contribution from the crossed
   one) for the two-gluon exchange DY amplitude which vanishes in the
   small $x$ limit.} 
\label{fig5dy}
\end{figure}
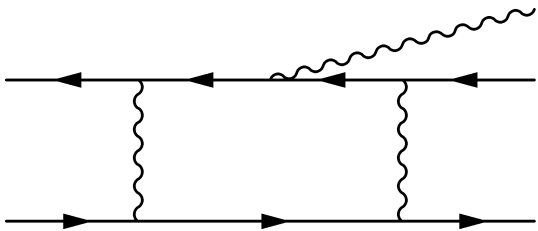

It is instructive to note the mathematical origin of this suppression,
as it occurs in the Feynman gauge calculation.
The diagram of Fig.~5 is suppressed because 
the poles in the (arbitrarily chosen) integration variable 
$k_2^+$ arising from the internal propagators $p_1+k_1$ and $p_2-k_2$ 
lie on the same half-plane\footnote{In $A^+=0$ gauge
the diagram of Fig.~5 can contribute, depending on the prescription
which is used to regularize the spurious $k_i^+=0$ pole of the 
gluon propagator in this gauge. The present discussion concerning
the location of {\it physical} poles holds in any gauge, but in  
$A^+=0$ gauge, a finite value for the diagram of Fig.~5 arises
when one uses for instance the principal value prescription, because this 
prescription involves spurious poles on both sides of the real axis.}.
Note that the associated Feynman gauge DIS diagram 
(obtained by the $q \leftrightarrow p_1$ exchange) is not suppressed
in the small $x_B$ limit because the corresponding propagators read
$p_1-k_1$ and $p_2-k_2$, yielding poles lying on different sides of
the real axis. Shifting the $p_1$ line from final (DIS) to initial (DY) 
state has non-trivial analytical consequences \cite{lrs}.

The result for the full DY one-loop amplitude is (compare to the 
one-loop DIS amplitude \eq{BDIS})
\beqa
\label{BDY}
B_{DY}(p_2^-,\pvec_{2\perp},\kvec_\perp) &=& - ieg^4 M Q p_2^- 
\left[\frac{1}{D(\pvec_{2\perp})} -
\frac{1}{D(\pvec_{2\perp}-\kvec_{\perp})} \right] 
\nonumber \\ 
&\times& \int \frac{d^2
\kvec_{1\perp}}{(2\pi)^2}\ \frac{1}{k_{1\perp}^2 k_{2\perp}^2}
\label{BexprDY}  
\eeqa
The infrared sensitivity of the amplitude $B_{DY}$ 
will be discussed below.

{\bf Three-gluon exchange}

Similarly to the one-loop case, radiation within the target 
is suppressed in the region \eq{MscalesDY}, where 
$k_{i\perp} \gg Mx$. In Feynman gauge only two diagrams 
(including obvious permutations) contribute
to the two-loop amplitude, corresponding to the three exchanges occurring
all before or all after the virtual photon emission. The result reads
(compare to the DIS amplitude \eq{CDIS})
\beqa
\label{CDY}
C_{DY}(p_2^-,\pvec_{2\perp},\kvec_\perp) &=& \frac{1}{3} eg^6 M Q
p_2^- \left[ \frac{1}{D(\pvec_{2\perp})}  -
\frac{1}{D(\pvec_{2\perp} -\kvec_{\perp})} \right] \nonumber \\
&\times& \int
\frac{d^2 \kvec_{1\perp}}{(2\pi)^2}
\frac{d^2 \kvec_{2\perp}}{(2\pi)^2}\ \frac{1}{k_{1\perp}^2\,
k_{2\perp}^2\, k_{3\perp}^2} 
\eeqa

\vskip 5mm
{\bf \large 3.3 Absence of DY shadowing for $k_{\perp} \gg Mx$}

We now discuss the expressions \eq{BDY} and \eq{CDY} for the
DY loop amplitudes.
Contrary to the case of DIS, they show an infrared sensitivity 
when $k_{i\perp} \to 0$. This infrared singularity 
is absent in the cross section, the Coulomb phase originating 
from scattering between charged particles cancelling between 
the production amplitude and its conjugate. However the {\it total}
DY cross section is out of reach within the present approximation
\eq{MscalesDY}. Indeed, the amplitudes have been evaluated with the
assumption $k_{i\perp}, k_{\perp} \gg Mx$, 
and their precise infrared behaviour can thus not be inferred.
However, as we will see now the {\it partial} contribution to the cross  
section originating from $k_{\perp} \gg Mx$ can be obtained. We will show
that this contribution actually vanishes (at order $e^2 g^8$).

For a finite $k_{\perp} \gg Mx$ the expression \eq{ADY} for the Born amplitude
is valid and \eq{BDY} and \eq{CDY} can be written as
\beqa
B_{DY} &=& i \frac{g^2}{2} A_{DY} \int 
\frac{d^2 \kvec_{1\perp}}{(2\pi)^2}\ 
\frac{k_{\perp}^2}{k_{1\perp}^2 k_{2\perp}^2} 
\label{BDY2}\\
C_{DY} &=&  - \frac{g^4}{6} A_{DY} \int
\frac{d^2 \kvec_{1\perp}}{(2\pi)^2}
\frac{d^2 \kvec_{2\perp}}{(2\pi)^2}\ \frac{k_{\perp}^2}{k_{1\perp}^2 k_{2\perp}^2 k_{3\perp}^2}
\label{CDY2}
\eeqa

One gets for the rescattering correction to the Born term:
\beqa
&& \frac{d \Delta \sigma_{DY}}{d^2\pvec_{2\perp}d^2\kvec_{\perp}} 
\propto |B_{DY}|^2 + 2 A_{DY} C_{DY} = 
|A_{DY}|^2 F(k_{\perp}^2) \label{DYcorrection} \\
&& F(k_{\perp}^2) \equiv \frac{g^4}{4(2\pi)^4} k_{\perp}^4 \left\{
    \left[R_2(k_{\perp})\right]^2 -\frac{4}{3} R_{13}(k_{\perp})  \right\}
\label{F}
\eeqa
where
\beqa
R_2(k_{\perp}) &=& \int \frac{d^2\kvec_{1\perp}}{k_{1\perp}^2(\kvec_\perp
-\kvec_{1\perp})^2} \nonumber \\
R_{13}(k_{\perp}) &=& \frac{1}{k_\perp^2} \int
\frac{d^2\kvec_{1\perp}\ d^2\kvec_{2\perp}}{k_{1\perp}^2
k_{2\perp}^2(\kvec_\perp -\kvec_{1\perp} -\kvec_{2\perp})^2}
\label{r2r13}
\eeqa

Let us stress that in the small $x$ limit, 
\eq{DYcorrection} is correct for any finite $k_{\perp}$, since 
in \eq{DYcorrection} also the momenta $k_{i\perp}$ flowing 
in the loops are large, $k_{i\perp} \gg Mx$. Indeed,
although the individual amplitudes are infrared singular, 
in dimensional regularization one obtains the non-trivial result  
(see for instance \cite{bhmps} where the same expression appeared
in another context):
\beq
\label{mastereq}
k_{\perp} \neq 0 \Rightarrow  F(k_{\perp}^2) \propto 
\left[R_2(k_{\perp})\right]^2 -\frac{4}{3} R_{13}(k_{\perp}) = 0
\eeq
The fact that $F(k_{\perp}^2)$ is 
infrared finite shows that the typical 
values of $k_{i\perp}$ in the loop integrals of \eq{r2r13}
are of order $k_{\perp}$, the only scale at disposal.
This justifies the approximation $k_{i\perp} \gg Mx$ used to evaluate
the loop amplitudes. 
But since moreover $F(k_{\perp}^2)=0$ for any finite $k_{\perp}$,
only small  $k_{\perp} \sim Mx \to 0$ may contribute to the
$k_{\perp}$-integrated correction $\Delta \sigma_{DY}$.

We obtain here the main result of this paper. For a fixed $k_{\perp}$
satisfying $k_{\perp} \gg Mx$, the rescattering correction (of relative
order $g^4$) to the DY Born cross section vanishes,
\beq
\label{kperpDY}
k_{\perp} \gg Mx \Rightarrow \frac{d\Delta \sigma_{DY}}{dk_{\perp}^2} =0 
\eeq
This is in contrast with the DIS situation, where 
$k_{\perp} \sim m_\pl \gg Mx$ contributes to the \order{g^4} correction  
to $\sigma_{DIS}$.
These features are similar to what was found by Bethe and Maximon for
high energy pair production and bremsstrahlung \cite{bm}.
At momentum transfers much larger than their minimal value, Coulomb
rescatterings modify the Born cross section for pair production, but not 
for bremsstrahlung. The absence of corrections to the parton model DY
formula was also found in a pre-QCD context \cite{cw,del}.

\vskip 1cm
{\bf \Large 4. Discussion}
\vskip 5mm

We showed within a simple abelian model that whereas for 
$k_{\perp} \sim m_\pl$ the DIS cross section gets a shadowing
correction, Coulomb rescatterings do not modify the DY Born 
cross section at similar $k_{\perp}$.

It is still possible that the {\it $k_{\perp}$-integrated} 
DY rescattering correction \eq{DYcorrection} could equal 
the result \eq{sigmaDIS} found in DIS, in agreement with universality.
But since the approximation \eq{MscalesDY} we used breaks down for 
$k_{\perp} \sim Mx$, we cannot integrate \eq{DYcorrection} down to 
such small $k_{\perp}$ va\-lues and thus cannot answer this question.
Calculating the DY amplitudes beyond the small $x$ limit would be much
more involved. In particular, for $k_{\perp} \sim Mx$ radiation in between
Coulomb scatterings is not suppressed.

Since $k_{\perp} \gg Mx$ induces unitary Coulomb corrections to the 
Drell-Yan cross section, any non-vanishing contribution of order $e^2g^8$ to 
$\Delta \sigma_{DY}$ must arise from the domain $k_{\perp} \sim Mx$, 
as stated in \eq{dydis}.
The fact that 
different $k_{\perp}$-ranges in DIS and DY could then sum up to identical
total cross sections for {\it any} target is not obvious.
In the case of a totally screened target, with inverse
screening length $\Lambda$, the values $k_{\perp} \sim Mx$ are
forbidden if $Mx \ll \Lambda$. One thus expects, for such values
of $x$, Coulomb rescatterings to affect the DIS cross
section but not the DY one (in the leading-twist regions of
interest). Relying on Eq.~\eq{dydis}, we thus 
suggest that the nucleon quark distribution functions probed
in DIS and DY might become non-universal when $M_N x \ll \Lambda$,
with $M_N$ the nucleon mass and $\Lambda \sim \Lambda_{QCD}$.
We roughly estimate that the violation of universality sets in when
the Ioffe time of the photon $\nu/Q^2 = 1/(2 M_N x)$ becomes larger 
than $1/\Lambda_{QCD}$, \ie\ when $x < 0.1$.

We found instructive to supplement 
our scalar QED model with a mass term for
the exchanged Coulomb photons. Calling $\lambda$ the photon mass, and 
considering the limit $\lambda \gg Mx$, the DY production amplitudes
in this modified model are simply obtained by the replacements
$k_{\perp}^2 \rightarrow k_{\perp}^2 + \lambda^2$ in \eq{ADY} and 
$k_{i\perp}^2 \rightarrow k_{i\perp}^2 + \lambda^2$ in \eq{BDY} and
\eq{CDY}. Then \eq{DYcorrection} can be integrated over the whole 
$k_{\perp}$-range, the photon mass $\lambda$ 
acting effectively as an infrared cutoff. We show in Appendix A
that in this specific case the integrated DY cross section 
$\Delta \sigma_{DY}$ is
identical (see \eq{sigmaDYunscreened}) to that of DIS given in \eq{sigmaDIS}.  
We also show that the typical value of $k_{\perp}$ is of order
$\lambda$. This illustrates that when
$k_{\perp}$ can reach its minimal value ($\lambda$ in the present case),
the two different $k_{\perp}$-ranges in DIS and DY {\it might} give equal
total contributions. Let us mention that a similar result was 
found by Bethe and Maximon for pair production and
bremsstrahlung, in the case of an unscreened target \cite{bm}.

This somewhat academic calculation may help
understanding why the universality of the quark 
distribution was claimed to hold in Refs.~\cite{bhq,kop}. In these papers
the DIS and DY cross sections depend on the same 
non-perturbative parameter (to be interpreted as the quark
distribution, see in particular \cite{bhq}),
namely the quark pair dipole cross section in the target, expressed
in impact parameter space.
It is what we find here (see the comments following 
Eq.~\eq{sigmaDYunscreened}), but in the very particular case of an
unscreened pointlike target and for a finite photon mass $\lambda \gg Mx$.
The fact that only small $k_{\perp} \sim \lambda$ contributes
(see \eq{mastereq2}) would appear difficult to infer in a
coordinate space approach. 
One indeed finds that the typical value of the impact parameter 
in \eq{sigmaDYunscreened} is
$\ave{R_\perp} \sim 1/m_\pl$. However the dominance of small 
$k_{\perp} \ll m_\pl$ for DY can be seen in our  
momentum space calculation, as expressed in \eq{mastereq2}. 
We explain in Appendix A why the relation 
$\ave{k_{\perp}}_{DY} \ll 1/\ave{R_{\perp}}$ is possible. 
(In particular we do not contradict the uncertainty principle.)
This point may have been overlooked in previous coordinate space 
approaches. We show in Appendix B that the derivation of the
color dipole formulation of the Drell-Yan process \cite{kop,bhq,rpn}
relies implicitly on the particular limit studied in Appendix A, 
namely $\lambda \gg Mx$. Apparently no general proof, valid in
the realistic limit $\lambda \to 0$ at fixed $Mx$, is known. 

The result \eq{dydis} is demonstrated in the present paper by
comparing leading-twist Coulomb rescattering corrections
in DIS and DY in a model 
with a pointlike target, but which however contains the relevant features 
of nuclear shadowing \cite{bhmps}. Our arguments indicate that
for a realistic target, leading-twist nuclear shadowing in DY might
be reduced compared to shadowing 
in DIS. The data on DIS \cite{nmc-1996,pw} and DY 
\cite{E772-1990,E866-1999} shadowing seem to be reasonably 
consistent with the assumption that nuclear leading-twist quark
distributions are universal, and any possible 
violation of universality can thus not be too large. But the
difficulty to disentangle valence and sea quark shadowing, as well as
quark energy loss effects \cite{arleo} makes phenomenological analyses 
particularly intricate. We think that a possible violation 
of universality at small $x$ is not 
ruled out by the existing data. 

\vskip 1cm

{\bf Acknowledgements.} I am most thankful to P.~Hoyer
for many discussions and advice during this work. I wish to thank also
F.~Arleo, S.~Brodsky, J.~Collins, D.~S.~Hwang and J.~Raufeisen 
for very helpful and instructive exchanges.

\eject

\vskip 1cm
\centerline{{\bf \Large Appendix A}}
\vskip 5mm
\centerline{{\bf \Large A particular limit: $\lambda \gg Mx$}}
\vskip 5mm

Here we show that in the particular case where the Coulomb photons are
given a finite mass $\lambda \gg Mx$, Coulomb rescatterings affect
identically the {\it total} DIS and DY cross sections, in agreement
with universality\footnote{We stress that the target, being a scalar
{\it charged} quark, is still unscreened. For a {\it neutral} target, 
the screening scale $\Lambda$ (intrinsic to the target form factor)
and the photon mass $\lambda$ would 
have a priori no reason to be identical.}. 
In this modified scalar QED model, the DY production amplitudes
\eq{ADY}, \eq{BDY} and \eq{CDY} are regularized in the infrared by 
$k_{i\perp}^2 \rightarrow k_{i\perp}^2 + \lambda^2$ (and denoted
by the subscript $\lambda$), and become in 
transverse coordinate space:
\beqa
\tilde A_{DY}^{\lambda}(p_2^-,\rvec_\perp, \Rvec_\perp) &=& -2eg^2 M Q p_2^-\,
V(m_\pl r_\perp) \left[ G(R_\perp) - G(|\Rvec_\perp+\rvec_\perp|) \right]
 \nonumber  \\
\tilde B_{DY}^{\lambda}(p_2^-,\rvec_\perp, \Rvec_\perp) &=& - ieg^4 M Q p_2^-\,
V(m_\pl r_\perp) \left[ G(R_\perp)^2 - G(|\Rvec_\perp+\rvec_\perp|)^2 \right]
\nonumber  \\
\tilde C_{DY}^{\lambda}(p_2^-,\rvec_\perp, \Rvec_\perp) &=& 
+ \frac{eg^6}{3} M Q p_2^-\,
V(m_\pl r_\perp) \left[ G(R_\perp)^3 - G(|\Rvec_\perp+\rvec_\perp|)^3 \right]
\nonumber \\ 
G(R_\perp) &\equiv&
\int \frac{d^2\kvec_\perp}{(2\pi)^2}
\frac{e^{i\Rvec_\perp\cdot\kvec_{\perp}}}{k_\perp^2 + \lambda^2 } 
= \frac{1}{2\pi} K_0(\lambda R_\perp)
\label{G} \nonumber \\
\label{ABCtildeDY}
\eeqa

Using the above expressions one obtains 
\beqa
\Delta \sigma_{DY} &=& 
\int d^2\rvec_\perp\, d^2\Rvec_\perp\,
\left[ |{\tilde B}_{DY}^{\lambda}|^2 
+ 2 {\tilde A}_{DY}^{\lambda} {\tilde C}_{DY}^{\lambda} \right] 
\nonumber \\
&=& - \frac{(eg^4 M Q p_2^-)^2 }{3}
\int d^2\rvec_\perp\, d^2\Rvec_\perp \, V(m_\pl r_\perp)^2 
\left[ G(R_\perp) - G(|\Rvec_\perp+\rvec_\perp|) \right]^4 \nonumber
\\
\label{lambdaDY1}
\eeqa
Assuming $\lambda \ll m_\pl$ the typical value 
of $R_\perp$ contributing to \eq{lambdaDY1} is
$\ave{R_\perp} \sim 1/m_\pl \sim \ave{r_\perp}$. Using  
$K_0(x) \simeq \log(1/x)$ for $x \ll 1$ we get:
\beq
\Delta \sigma_{DY} = - \frac{1}{3} \int d^2\rvec_\perp\, d^2\Rvec_\perp \, 
\frac{g^4}{4} {\tilde A}^2 W^2 
\label{sigmaDYunscreened}
\eeq
where ${\tilde A}$ and $W$ are given in Eqs.~\eq{AtildeDIS} and \eq{Wexpr}.
Comparing \eq{sigmaDYunscreened} to \eq{sigmaDIS} one sees that in
this particular case (finite photon mass 
$\lambda \gg Mx$) the leading-twist Coulomb corrections to the {\it total}
DIS and DY cross sections are identical.
Note also that similarly to \cite{bhq,kop}, the same quantity $W^2$ 
(the quark pair dipole rescattering cross section) appears in both DIS
and DY results. 

We now show that whereas $\ave{k_{\perp}}_{DIS} \sim m_\pl$
contributes to the DIS cross section \eq{sigmaDIS}, 
$\ave{k_{\perp}}_{DY} \sim \lambda \ll m_\pl$ contributes to 
\eq{sigmaDYunscreened}. The equations 
\eq{DYcorrection}, \eq{F} and \eq{r2r13}
are modified according to 
\beqa
&& \frac{d \Delta \sigma_{DY}}{d^2\pvec_{2\perp}d^2\kvec_{\perp}} 
\propto |B_{DY}^{\lambda}|^2 + 2 A_{DY}^{\lambda} C_{DY}^{\lambda} = 
|A_{DY}^{\lambda}|^2 F^{\lambda}(k_{\perp}^2) \label{lambdaDYcorrection} \\
&& F^{\lambda}(k_{\perp}^2) \equiv \frac{g^4}{4(2\pi)^4} 
(k_{\perp}^2 + \lambda^2)^2 \left\{
    \left[R_2^{\lambda}(k_{\perp})\right]^2 
-\frac{4}{3} R_{13}^{\lambda}(k_{\perp})  \right\}
\label{lambdaF}
\eeqa
where
\beqa
R_2^{\lambda}(k_{\perp}) &=& \int 
\frac{d^2\kvec_{1\perp}}{(k_{1\perp}^2 + \lambda^2)
  ((\kvec_\perp
-\kvec_{1\perp})^2 + \lambda^2) } \nonumber \\
R_{13}^{\lambda}(k_{\perp}) &=& \frac{1}{(k_\perp^2+ \lambda^2)} \int
\frac{d^2\kvec_{1\perp}\ d^2\kvec_{2\perp}}{(k_{1\perp}^2+ \lambda^2)
(k_{2\perp}^2+ \lambda^2)((\kvec_\perp -\kvec_{1\perp}
-\kvec_{2\perp})^2+ \lambda^2)} \nonumber \\
\label{lambdar2r13}
\eeqa

For $k_{\perp} \gg \lambda$ a difficult calculation yields:
\beq
\label{mastereq2}
k_{\perp} \gg \lambda \Rightarrow   
\left[R_2^{\lambda}(k_{\perp})\right]^2 
-\frac{4}{3} R_{13}^{\lambda}(k_{\perp}) 
\sim \morder{\frac{\lambda^2}{k_{\perp}^6}\log^2
\left(\frac{k_{\perp}}{\lambda}\right)}
\eeq
This latter equation expresses within a mass regularization scheme the
result \eq{mastereq} obtained in dimensional regularization.
Thus for $\lambda \ll k_{\perp} \ll p_{2\perp}$, the integrand 
\eq{lambdaDYcorrection} behaves as $\sim \lambda^2/k_{\perp}^4$ (since 
$A_{DY}^{\lambda} \propto \kvec_{\perp}/k_{\perp}^2$ in this range)
and the contribution from $k_{\perp} \gg \lambda$ is suppressed in the 
$k_{\perp}$-integrated quantity $\Delta \sigma_{DY}$. We conclude that 
$\ave{k_{\perp}} \sim \lambda \ll m_\pl$ dominates in 
$\Delta \sigma_{DY}$. In the present particular case 
($\lambda \neq 0$), this is reminiscent from the result 
$\ave{k_{\perp}}_{DY} \sim Mx \ll m_\pl$ we derived in section 3.

The fact that $\ave{R_{\perp}} \sim 1/m_\pl$ and 
$\ave{k_{\perp}} \sim \lambda \ll m_\pl$ does not contradict the 
uncertainty principle. One can easily see that the one-loop amplitude
$B_{DY}^{\lambda}$ appearing in \eq{lambdaDYcorrection} behaves as 
$(\kvec_{\perp}/k_{\perp}^2) \log(k_{\perp}^2/\lambda^2)$ for 
$\lambda \ll k_{\perp} \ll m_\pl$. Hence 
$\int d^2\kvec_{\perp} |B_{DY}^{\lambda}(\kvec_{\perp})|^2$ is
dominated by the {\it logarithmic} range 
$\lambda \ll k_{\perp} \ll m_\pl$. In impact parameter space the
expression 
$\int d^2\vec{R}_{\perp}
|\tilde{B}_{DY}^{\lambda}(\vec{R}_{\perp})|^2$ is dominated by the 
logarithmic interval $1/m_\pl \ll R_{\perp} \ll 1/\lambda$, as can be
seen from the expression of $\tilde{B}_{DY}^{\lambda}$ given in 
\eq{ABCtildeDY}, and as expected from the uncertainty principle. A
similar conclusion is obtained for the term 
$\sim A_{DY}^{\lambda} C_{DY}^{\lambda}$ in \eq{lambdaDYcorrection}.
However, performing the sum $|B|^2 + 2AC$ suppresses the regions 
$k_{\perp} \gg \lambda$ in momentum space (see \eq{mastereq2}), and 
$R_{\perp} \gg 1/m_\pl$ in coordinate space (see \eq{lambdaDY1}
and \eq{sigmaDYunscreened}). As a result 
$\ave{k_{\perp}} \sim \lambda$ and $\ave{R_{\perp}} \sim 1/m_\pl$.
This does not contradict the uncertainty principle because the
function $\sqrt{|\tilde{B}|^2 + 2\tilde{A}\tilde{C}}$ 
is obviously not the Fourier transform of $\sqrt{|B|^2 + 2AC}$.
We note that $\ave{k_{\perp}} \ll 1/\ave{R_{\perp}}$ is possible
thanks to the presence of two different scales, $m_\pl$ and
$\lambda$, and to the logarithmic spread in the separate contributions 
from $|B|^2$ and $2AC$ to the DY cross section. This feature is absent in
DIS. The DIS amplitudes are not infrared sensitive and thus only the scale 
$m_\pl$ is relevant.

\vskip 1cm
\centerline{{\bf \Large Appendix B}}
\vskip 5mm
\centerline{{\bf \Large Dipole formulation of the Drell-Yan process}}
\vskip 5mm

In this Appendix we discuss more precisely the correspondence between
our model calculation and the dipole formulation of the Drell-Yan
process \cite{kop,bhq,rpn}. We first note that this formulation was
originally proposed in Ref.~\cite{kop}, on the basis of a {\it Born}
calculation. The Born DY diagrams obtained from Figs.~2a and 2b by the
$q \leftrightarrow p_1$ exchange were calculated in impact parameter
space in \cite{kop}. In the present scalar QED model the Fourier
transform of \eq{ADY} reads
\beq
\tilde A_{DY}(p_2^-,\rvec_\perp, \Rvec_\perp) 
= - 2eg^2 M Q p_2^-\, V(m_\pl r_\perp) W(\rvec_\perp, \Rvec_\perp)
\label{AtildeDY}
\eeq
The Born DY cross section is thus:
\beq
\label{BornDY}
\sigma_{DY}^{Born} \sim
\int d^2\rvec_\perp\, d^2\Rvec_\perp\,
|{\tilde A}_{DY}|^2 =  4 (eg^2 M Q p_2^-)^2 
\int d^2\rvec_\perp\, d^2\Rvec_\perp \, V^2 W^2
\eeq
At the Born level, the DIS and DY cross sections are 
identical because the DIS and DY Born amplitudes given
in \eq{ADIS} and \eq{ADY} are the same (up to a sign). Thus the DIS
quark pair dipole scattering cross section $W^2$ appears in \eq{BornDY}.
We thus recover in a simple framework the dipole
formulation (obtained at the Born level) of the Drell-Yan process 
proposed in \cite{kop}. 

In order to see what happens at higher orders, we sum over any number
of Coulomb rescatterings. An obvious generalization of \eq{BDY} and
\eq{CDY} yields:
\beqa
{\cal M}_{DY} &=&  A_{DY} + B_{DY} + C_{DY} + \ldots  \\
&=& 2 ie  M Q p_2^- \left[\frac{1}{D(\pvec_{2\perp})} -
\frac{1}{D(\pvec_{2\perp}-\kvec_{\perp})} \right] {\Delta}(\kvec_\perp)
\label{MDY}
\\ \nonumber \\ 
\Delta(\kvec_\perp) &=& \frac{ig^2}{k_{\perp}^2} - \frac{g^4}{2!} 
\int \frac{d^2
\kvec_{1\perp}}{(2\pi)^2}\ \frac{1}{k_{1\perp}^2 k_{2\perp}^2}
- \frac{ig^6}{3!} \int
\frac{d^2 \kvec_{1\perp}}{(2\pi)^2}
\frac{d^2 \kvec_{2\perp}}{(2\pi)^2}\ \frac{1}{k_{1\perp}^2\,
k_{2\perp}^2\, k_{3\perp}^2} + \ldots  \nonumber \\ 
\label{delta}
\eeqa 
where $\kvec_{1\perp}+ \ldots + \kvec_{n\perp}= \kvec_{\perp}$ in the
denominators $k_{1\perp}^2 \ldots k_{n\perp}^2$ appearing in
\eq{delta}. As already discussed in section 3.3, the expression 
\eq{delta} arises from the assumption $k_{\perp},\, k_{i\perp} \gg Mx$
used to evaluate the Born and loop amplitudes, and contains infrared
singularities when $k_{i\perp} \to 0$. Thus integrating over
$\kvec_\perp$, as well as Fourier transforming \eq{MDY} to impact
parameter space cannot be done, since the physical infrared regulators
$\sim Mx$ have been neglected. Regularizing \eq{delta} by 
$k_{i\perp}^2 \rightarrow k_{i\perp}^2 + \lambda^2$ with the 
{\it same} parameter $\lambda$ in all de\-no\-mi\-na\-tors 
amounts to implicitly assume $\lambda \gg Mx$, as was done in Appendix
A. After this regularization is done in \eq{delta}, Fourier
transforming \eq{MDY} gives:
\beqa
\label{MtildeDY}
{\tilde \M}_{DY}(p_2^-,\rvec_\perp, \Rvec_\perp) 
&=& 2 i e M Q p_2^-\, V(m_\pl r_\perp) \nonumber \\ 
&\times& 
\left[ \exp(ig^2 G(R_\perp))-  \exp(ig^2 G(|\Rvec_\perp+\rvec_\perp|))
\right]
\eeqa
where $G(R_\perp)$ is given in \eq{ABCtildeDY}. The equation
\eq{MtildeDY} is similar to what was obtained in \cite{bhq,rpn}. The
first and second terms arise from photon radiation respectively after
and before the Coulomb exchanges, and involve quark impact parameters
which are shifted by the amount $\rvec_\perp$. We stress here that
these terms are infrared singular when $\lambda \to 0$. Squaring
\eq{MtildeDY} one gets:
\beqa
\label{MtildeDYsquared}
\int d^2\rvec_\perp\, d^2\Rvec_\perp\,
|{\tilde {\cal M}}_{DY}|^2 &=& (4e M Q p_2^-)^2 \int d^2\rvec_\perp\,
d^2\Rvec_\perp\, V(m_\pl r_\perp)^2 \nonumber \\ 
 &\times& \sin^2\left[ \frac{g^2}{2}
  (G(R_\perp)-G(|\Rvec_\perp+\rvec_\perp|)) \right] 
\eeqa
This is infrared finite when $\lambda \to 0$, and equals the result
obtained in DIS to all orders in $g$ \cite{bhmps}: 
\beq
\label{limitMtildeDYsquared}
\int d^2\rvec_\perp\, d^2\Rvec_\perp\,
|{\tilde {\cal M}}_{DY}|^2 
\ \ \mathop{\longrightarrow}_{\lambda \to 0} \ \ 
\int d^2\rvec_\perp\, d^2\Rvec_\perp\, |{\tilde A}_{DY}|^2 
\left[ \frac{\sin(g^2W/2)}{(g^2W/2)}   \right]^2
\eeq
In particular the term of order $g^8$ in the expansion of 
\eq{limitMtildeDYsquared} reproduces \eq{sigmaDIS} (and
\eq{sigmaDYunscreened}). The identity of the DY and DIS cross sections
found here arises from first assuming $\lambda \gg Mx$, and then
taking the $\lambda \to 0$ limit. This procedure was implicitly used
in \cite{bhq,rpn}. In these works the step consisting in the
replacement
$\displaystyle{G(R_\perp)-G(|\Rvec_\perp+\rvec_\perp|)
\mathop{\longrightarrow}_{\lambda \to 0} W(\rvec_\perp, \Rvec_\perp)}$  
(see \eq{MtildeDYsquared} and \eq{limitMtildeDYsquared}) is made
without mentioning that $G(R_\perp)$ is a well-defined quantity only
in the presence of $\lambda \neq 0$.
Thus the dipole formulation of the DY process \cite{kop,bhq,rpn} is in
fact established beyond the Born approximation within the limit 
$\lambda \gg Mx$. Independently of the question whether this 
formulation holds also in the limit 
$\lambda \to 0$ at fixed $Mx$, we stress that the very expression
\eq{MtildeDY} is obtained from an integration over {\it all} 
$\kvec_\perp$'s down to very small values $\sim \lambda$, as shown in
Appendix A. That such an expression can be obtained in general, for
any realistic neutral target, has to our knowledge not been proven.


\end{fmffile}
\end{document}